\documentclass[%
 reprint,
 amsmath,amssymb,
 aps,
]{revtex4-2}

\usepackage{comment}
\usepackage[normalem]{ulem}
\usepackage{graphicx}% Include figure files
\usepackage{dcolumn}% Align table columns on decimal point
\usepackage{bm}% bold math
\usepackage{hyperref}% add hypertext capabilities
\usepackage{soul,xcolor}
\setstcolor{red}
\newcommand*{\rom}[1]{\expandafter\romannumeral #1}
\newcommand{\brk}[1]{\left( #1  \right)}
\newcommand{\abs}[1]{\left| #1  \right|}

\begin{document}

\title{Far from equilibrium relaxation in the weak coupling limit}% Force line breaks with \\
%\thanks{A footnote to the article title}%
% 
\author{Ran Yaacoby}
\affiliation{%
Department of Condensed Matter Physics, Tel Aviv University, Tel Aviv 69978, Israel
}
\author{Oren Raz}
\affiliation{%
 Department of Physics of Complex Systems, Weizmann Institute of Science, Rehovot 7610001, Israel
}%
 % \email{oren.raz@weizmann.ac.il}
\author{Gianluca Teza}
\email{teza@pks.mpg.de}
\affiliation{Max Planck Institute for the Physics of Complex Systems, Nöthnitzer Str. 38, 01187 Dresden, Germany}

\date{\today}

\begin{abstract}
It is commonly assumed that a large system, weakly coupled to a thermal environment through its boundaries, relaxes quasistatically towards the new equilibrium even when the temperature of the environment changes abruptly.  Here we show how this intuitive picture can break down for discrete energy systems, even in the case of infinitely weak coupling.
We provide an example in the Ising chain, showing how the interaction among degrees of freedom can create corrugated energy landscapes that are responsible for far-from-equilibrium and allow anomalous relaxation effects to survive infinitely weak couplings.
\end{abstract}

\maketitle

A physical system is characterized by its degrees of freedom (DoF), representing independent variables that change with time as the system evolves.
All of the combinations that the DoF can acquire define the system's possible microscopic configurations -- or microstates -- which is infinite for continuous DoF and finite in the discrete case \cite{Callen1985}.
In isolated systems, the total energy is preserved, and the microstate changes only between those corresponding to the same energy. Coupling the system to a thermal bath enables transitions between microstates with different energy, and the energy difference is compensated by heat exchange with the bath. The system then evolves towards thermal equilibrium with the bath, with the microstates' distribution approaching the Boltzmann distribution at bath temperature $T_b$ \cite{Callen1985}.

In macroscopic systems the coupling with the bath is commonly limited to the system boundaries, which consists of a considerably smaller (sub-extensive) portion of the DoF. In these cases, the boundary DoF are able to exchange heat with the bath \cite{Teza2021_mpemba_boundary}, while the rest of the DoF exchange energy with the boundary DoF and among themselves in an \emph{ergodization} process, which redistributes the energy within the system, while keeping the total energy fixed.
The combination of heat exchange through the boundaries and internal ergodization can generate strikingly counterintuitive phenomena like the Mpemba \cite{Lu2017,Klich2019,Gal2020,Lasanta2017,Kumar2020,Jeng2006,chaddah2010overtaking,Hu2018,Kumar2020,bechhoefer2021fresh,walker2021,kumar2022anomalous,Teza2021_mpemba_boundary,pemartin2024shortcuts,walker2022mpemba,bera2023effect,walker2023optimal} and Kovacs \cite{Kovacs1964,Bertin2003,Mossa2004,prados2014kovacs,bouchbinder2010nonequilibrium,ruiz2014kovacs,lahini2017nonmonotonic} effects, dynamical phase transitions \cite{teza2022eigenvalue} and the cooling/heating asymmetry \cite{lapolla2020faster,ibanez2024heating}.
Anomalous relaxation effects have been predicted and verified also in open quantum systems \cite{Carollo2021,chatterjee2023quantum,aharony2024inverse} with non-thermal analogs finding application in closed Hamiltonian systems \cite{ares2023entanglement,rylands2024microscopic,joshi2024observing}.
These effects share a common feature: the system explores far-from-equilibrium distributions during the relaxation process.
Intuitively, this is not expected in the weak coupling limit, where the slow timescale of the heat exchange with the bath should allow the system to thermalize as soon as some energy is exchanged, implying a quasistatic relaxation across the entire evolution.

In this Letter, we show that in contrast to the intuitive picture above, a system that is initiated in equilibrium with a bath at a temperature $T_0$ can relax upon a quench to a bath with temperature $T_b$ through a path that is far from \emph{any} equilibrium distribution, regardless of the coupling strength.
This holds true even in the weak coupling limit, with infinite time separation between ergodization and heat exchange rates. 
We attribute this counterintuitive phenomenon to two main factors: discrete energy values and multiple macrostate variables.
We show how these factors force the relaxation trajectory to deviate from equilibrium distributions, and how a wide class of standard out-of-equilibrium dynamics induce a trajectory that surprisingly also differs from the gradient descent of the Landau free energy \cite{jordan_free_1997,kardar2007model}.
We provide an example in a paradigmatic many-body interacting model exhibiting the above phenomenology -- the 1D Ising antiferromagnet. These results are valid regardless of the exact details of the ergodization dynamic, since its details are lost in the fast ergodization limit \cite{Teza2021_mpemba_boundary}.

%%%%%%%%%%%%%%%%%%%%%%%%%%%%%%%%%%%%%%%%%%%%%%%%%%%%%%%

\begin{figure}
    \centering
    \includegraphics[width=0.99\linewidth]{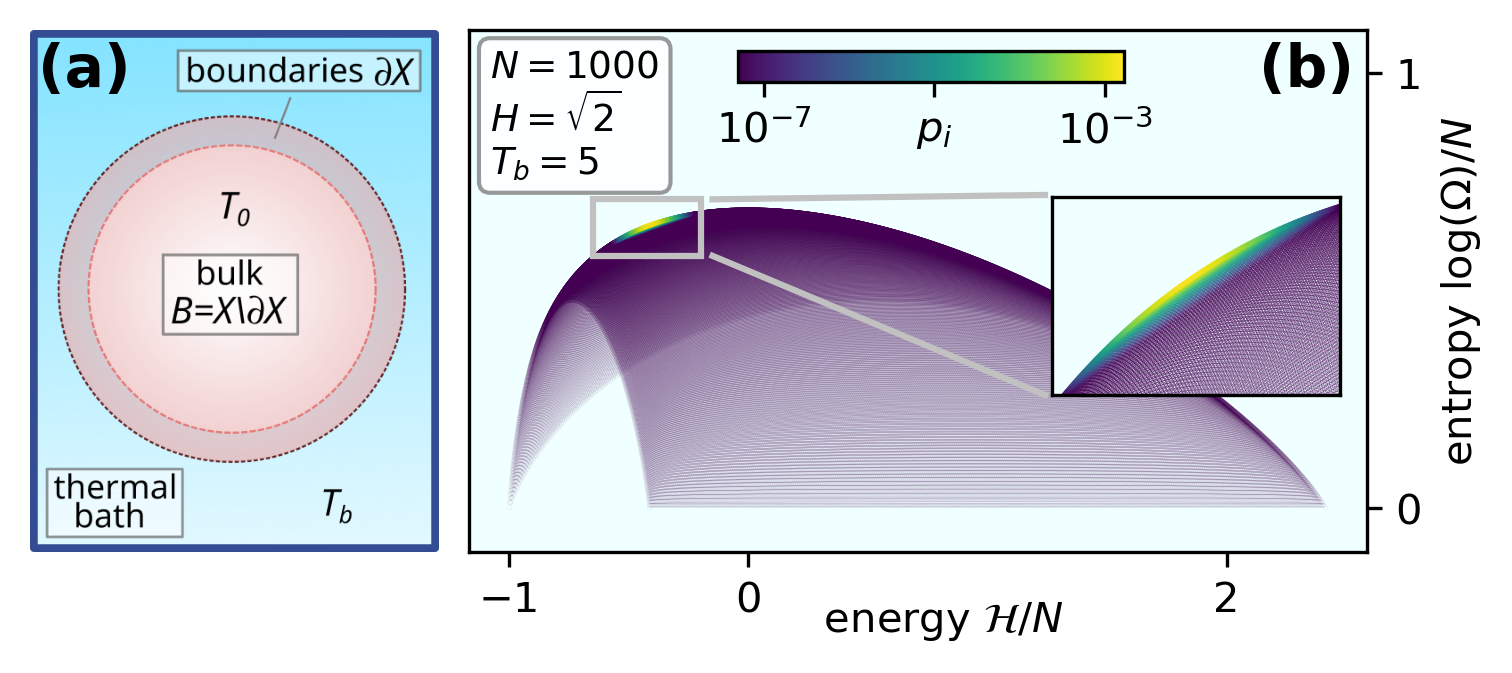}
    \caption{(a) Setup: the system exchanges heat with the bath only through the boundaries, while the bulk is completely isolated from the environment.
    (b) Representation of the energy macrostates of an  antiferromagnetic 1D Ising chain with $N=1000$ spins, plotted against their energy and entropy (each colored point correspond to an energy macrostate).
    The color of each energy macrostate corresponds to its Boltzmann equilibrium weight $\vec{\pi}(T_b)$. It is concentrated around the maximal entropy for a given energy (inset).}
    \label{fig:sketch}
\end{figure}

Limiting our discussion to finite discrete systems for simplicity, we refer to $X\equiv\{x_1,x_2,\dots,x_N\} $ as the set of $N$ DoF composing the system, and to a specific subset of these, $\partial X \subset X$, as the boundary DoF. 
The bulk of the system is $B\equiv X \setminus \partial X$ (Fig. \ref{fig:sketch}a).
The state of all $N$ DoF defines the microstate of the system.
The weak coupling limit corresponds to the limit of infinitely fast ergodization, so we can aggregate all microscopic configurations belonging to the same energy macrostate (i.e., sharing the same total energy) and construct an effective macroscopic dynamics that accounts for all the underlying microscopic transitions \cite{Teza2021_mpemba_boundary,Teza2020,Teza2020b}.
This procedure allows to describe the system's evolution with the master equation
\begin{equation}\label{eq:mast_eq}
    \partial_t \vec{p}(t)=\mathbf{W}^{\text{weak}}(T_b) \vec{p}(t)
\end{equation}
where $p_i(t)$ is the probability of the system to be in the energy-macrostate $i$ at time $t$, $\mathbf{W}^{\text{weak}}_{ij}$ is the transition rate from the energy-macrostate $j$ to the energy-macrostate $i$, and the diagonal terms $\mathbf{W}^{\text{weak}}_{ii}=-\sum_{j\neq i}\mathbf{W}^{\text{weak}}_{ji}$ are the escape rates from the energy-macrostate $i$.  
We denote the energy of the macrostate $i$ by $E_i$, and its multiplicity by $\Omega_i$. 
Assuming ergodicity and detailed balance, when the system is coupled to a bath with temperature $T_b$ it relaxes in time towards the Boltzmann distribution $\pi_i(T_b)=\Omega_ie^{-E_i/T_b}/\mathcal{Z}(T_b) \equiv e^{-\mathcal{F}_i/T_b}/\mathcal{Z}(T_b)$ where $\mathcal{Z}(T_b)=\sum_i e^{- \mathcal{F}_i/T_b}$ is the partition function at bath temperature, and we introduced the Landau free energy (LFE)
\begin{equation} \label{eq:landau_free_energy}
    \mathcal{F}_i = E_i - T_b \log\Omega_i,
\end{equation} 
with $k_B=1$.
Integrating Eq. \eqref{eq:mast_eq} we obtain
\begin{equation}\label{eq:mast_eq_solution}
    \vec{p}(t,T_0,T_b) = e^{\mathbf{W}^{\text{weak}}(T_b)t}\vec{\pi}(T_0)
\end{equation}
where we chose for initial condition at $t=0$ a Boltzmann equilibrium $\vec{\pi}(T_0)$ for some initial temperature $T_0$.
Starting from a Boltzmann equilibrium, we are able to exclude that the out-of-equilibrium phenomenology we want to address is an artifact due to a peculiarity of the initial conditions, proving that it directly stems from intrinsic properties of the system and the relaxation dynamics.

Using Eq. \ref{eq:mast_eq_solution} allows us to directly focus on the energy-macrostates to characterize and visualize the relaxation trajectory in the weak coupling limit. We introduce the set of $M$ macroscopic order parameters that uniquely define the energy macrostate of the system, $m_1, m_2, \dots, m_M$. In other words, an energy macrostate is defined through $\vec{m} \equiv (m_1, m_2, \dots, m_M )$.
As an example, in an Ising antiferromagnet one can choose the total magnetization and domain walls number (or, equivalently, the staggered magnetization) as two relevant observables \cite{Klich2019,teza2022eigenvalue,Teza2021_mpemba_boundary}.
We express the energy of the macrostate $\vec m_i$ by $E_i = E \brk{ \vec{m}_i }$ and its multiplicity by $\Omega_i = \Omega\brk{ \vec{m}_i }$, and consequently the LFE as $\mathcal{F}_i = \mathcal{F} \brk{ \vec{m}_i}$.
When the system is composed by discrete DoF, one generally has that
\begin{equation}\label{eq:assumption}
    E \brk{ \vec{m}_i } \neq E \brk{ \vec{m}_j } \quad \text{for any} \quad  i\neq j,
\end{equation}
implying that the energy uniquely defines the corresponding energy-macrostate $\vec m$.
For an Ising antiferromagnet, this is the case whenever the external magnetic field $H$ and coupling $J$ are incommensurate \cite{Klich2019,Vives1997}.
Note that Eq. \eqref{eq:assumption} doesn't hold in general for continuous DoF, where the equation $E(\vec m)=\textrm{const.}$ is in general an $M-1$ dimensional surface in the set of possible order parameters.   

Assuming Eq. \eqref{eq:assumption}, it is possible to identify the probability of being in a given energy macrostate as $p_i \brk{ t } \equiv p \brk{ \vec{m}_i, t }$ and to write the transition rate matrix as $\mathbf{W}^{\text{weak}}_{ij} \equiv \mathbf{W}^{\text{weak}} \brk{ \vec{m}_i, \vec{m}_j}$. This transforms the master Eq. \eqref{eq:mast_eq} into
\begin{equation}\label{eq:macro_master_eq}
    \begin{split}
        \partial_t p \brk{ \vec{m}_i, t } = \sum_{j} [ & \mathbf{W}^{\text{weak}} \brk{\vec{m}_i,  \vec{m}_j} p \brk{ \vec{m}_j , t } \\
        - & \mathbf{W}^{\text{weak}}  \brk{\vec{m}_j,  \vec{m}_i } p \brk{ \vec{m}_i, t } ]\
    \end{split}
\end{equation}
where we dropped the explicit dependencies on temperatures to simplify the notation.
As the system obeys detailed balance, the transition rates can be expressed as
\begin{equation} \label{eq:rates_decompose}
    \mathbf{W}^{\text{weak}} \brk{\vec{m}_i,  \vec{m}_j} = \Gamma_{ij} e^{ -\frac{1}{2 T_b} \brk{\mathcal{F}_j - \mathcal{F}_i}},
\end{equation}
where $\Gamma_{ij} \equiv \Gamma  \brk{ \mathcal{F}_i ,  \mathcal{F}_j }$ is a dynamics-dependent symmetric matrix.
In the most common dynamics algorithms e.g. Metropolis \cite{whitlock1986monte}, Glauber \cite{Glauber1963,Felderhof1971}, heat-bath \cite{newman1999monte,binder2012monte,teza2020rate} and Arrhenius \cite{mandal_proof_2011}, the symmetric part of the rates depends only on the difference in the LFE between the two states, namely $\Gamma_{ij} = g  \left(  (\mathcal{F}_j - \mathcal{F}_i)/T_b \right)$ for some even function $g$.
In Appendix A we report an analogous derivation for the most general case without assuming this parameterization.

The transitions from a generic macrostate $\vec{m}_i$ are restricted to some $\vec{m}_j = \vec{m}_i + \vec{\Delta}_{\alpha}$ where $\vec{\Delta}_{\alpha}$ is an $M$-dimensional difference vector connecting two macrostates according to the original microscopic transitions (Fig. \ref{fig:KL_en_fract}c).
These difference vectors are not scaling with the system size (in an Ising system, $\vec \Delta_\alpha $ describes the possible changes in the total magnetization and domain wall number due to a single spin-flip), in contrast with the macrostates $\vec{m}$ which are extensive.
Therefore, in the thermodynamic limit we can treat all the functions in Eq. \eqref{eq:macro_master_eq} as continuous functions of $\vec{m}$ and obtain the weak coupling limit relaxation dynamics (see Appendix A)
\begin{equation} \label{eq:me_continuum}
    \begin{aligned}
        \partial_t p \brk{ \vec{m}, t }
        = 
        \sum_{\alpha} \nabla_{\alpha} 
        [ & g \brk{ \frac{\nabla_\alpha \mathcal{F}  \brk{ \vec{m}}}{T_b}} \\
        &\times \sinh \brk{ \frac{\nabla_\alpha \mathcal{F} \brk{ \vec{m}} }{2 T_b} }  p \brk{ \vec{m}, t }],
    \end{aligned}
\end{equation}
where $\nabla_{\alpha}  = \sum_n \Delta_{n;\alpha} \partial_{m_n}$ is the derivative in the direction of $\vec{\Delta}_{\alpha}$.
The summation goes over all the directional derivatives, corresponding to the nearest neighbors $\alpha$ of $\vec{m}$.
Note that the second order terms scale like $\sim N^{-2}$ which, compared to $\nabla_{\alpha}$, become negligible in the thermodynamic limit.

Eq. \eqref{eq:me_continuum} displays two features that clearly explain why the relaxation process deviates from a quasistatic trajectory: i) all the terms in Eq. \eqref{eq:me_continuum} are explicit functions of the LFE at bath temperature $T_b$; and ii) the direction of the relaxation is dictated by the connectivity among macrostates, which is a specific property of the dynamics -- not the equilibrium.
In contrast, a quasistatic relaxation would require the dynamics to follow at any time $t>0$ the gradient of the Boltzmann equilibria with respect to temperature, independently of the LFE at $T_b$ and the connectivity.
This kind of scenario can be recovered when the system is described by a single order parameter ($M=1$), in which all evolutions occur along the same direction which is that of quasistatic trajectories.

Another result stemming from Eq. \eqref{eq:me_continuum} is that the relaxation process deviates from the gradient descent of the LFE, and does not follow the quasistatic trajectory either. This behavior contrasts with known models \cite{jordan_free_1997, kardar2007model}, in which the dynamics is naturally induced by the landscape of the free energy. In Eq. \eqref{eq:me_continuum}, on the other hand, one can clearly see how the freedom in choosing $g(\Delta F/T_b)$ affects the relaxation trajectory in probability space.
In Appendix A we provide two relevant examples (Glauber and Metropolis dynamics) in which the difference from LFE gradient descent is demonstrated, as well as the limit that retrieve the LFE gradient decent relaxation.

Let us present a paradigmatic example to illustrate the far from equilibrium relaxation in the weak coupling limit: the 1-dimensional Ising antiferromanget.
The $N$ spins $\sigma_s=\pm1$ define the DoF of the system $X=\{\sigma_1,\dots,\sigma_N\}$ and the $2^N$ corresponding microstates.
The Hamiltonian for any configuration $\vec{\sigma}\in X$ is
\begin{equation}\label{eq:ham_AF}
    \mathcal{H}(\vec{\sigma})= - H\sum_{s=1}^{N} \sigma_{s} - J\sum_{ s=1 }^{N} \sigma_{s} \sigma_{s+1} \equiv  - H S_1 -J S_2,
\end{equation}
where $J$ is the coupling constant, $H$ is an external magnetic field and $\sigma_{N+1}\equiv \sigma_1$. 
Here, we introduced two order parameters $\vec{m} = (S_1, S_2)$: the magnetization $S_1 = \sum_s \sigma_{s}$ and the domain walls number $S_2 = \sum_s \sigma_{s} \sigma_{s+1}$, both taking integer values in the range $[-N, N]$ with step sizes 2 and 4, respectively.
Here we choose $J=-1$ (antiferromagnetic interaction) for a clearer visualization of the phenomenology, which can nevertheless be observed in ferromagnets (see Fig. \ref{fig:S1_S2_ferro} in the End Matter).
We set $H=\sqrt{2}$ to ensure incommensurability with respect to $J$.
As Eq. \eqref{eq:assumption} holds, we identify the energy macrostates with the systems macrostates $\vec{m}_i = (S_{1;i}, S_{2;i})$. The number of microconfigurations grows exponentially with $N$, while energy macrostates grow quadratically as $(N / 2)^2+2$ \cite{Teza2021_mpemba_boundary,Teza2020,Teza2020b}.
This exact coarse-graining considerably extends the maximum system size that can be analyzed numerically.

In the boundary coupling setup, only a subset $\partial X\subset X$ of the DoF are connected to the bath. 
We assume that, apart from boundary flips, the system is completely isolated and therefore any additional dynamics cannot change the total energy in the system.
Consequently, it is limited to transitions between microstates within the same energy macrostate. 
Moreover, in the weak coupling limit the relaxation timescale within an energy macrosstates is extremely fast compared to the heat exchange, so we can consider the distribution within each energy macrostate uniformly ergodized.
For concreteness, we consider the case of a single spin coupled to the bath, $\partial X =\{\sigma_1\}$.
A general energy macrostate $\vec{m}_i$ is connected through thermal flips to other energy macrostates $\vec{m}_j$ only when $S_{1;i} - S_{1;j} = \pm 2$ and $S_{2;i} - S_{2;j} = \{0 , \pm 4\}$, which are the possible values for flipping one spin in a ring \footnote{The single-spin flip analysis implies similar analyses for larger sets $\partial X$.}. 
Using Glauber dynamics \cite{Glauber1963,Felderhof1971} one can parametrize the transition rates among these energy macrostates as
\begin{equation}\label{eq:rate_mat_weak}
    \mathbf{W}^{\text{weak}}_{ij}= \Gamma_{ij}e^{ \frac{ \mathcal{F} \brk{ \vec{m}_j, T_b } - \mathcal{F} \brk{ \vec{m}_i, T_b } }{2 T_b} },
\end{equation}
where $\Gamma_{ij} = \frac{1}{2}\textrm{sech } \frac{E(\vec{m}_j) - E(\vec{m}_i) }{2 T_b}$.
To find the LFE of each macrostate $\vec{m}_i$, we evaluate the exact multiplicity of states $\Omega_i$ combinatorially \cite{Andriushchenko2016}.
In more complicated models, like 2D Ising lattices \cite{Beale1996}, the exact form of $\Omega$ as a function of $E$ becomes quickly intractable as system size increases \footnote{Monte Carlo algorithms that estimate this dependence \cite{Wang2001} allow to explore relatively large sizes, but the approximations introduced could alter the relaxation mechanisms hindering the thermodynamic limit analysis.}.

In Fig. \ref{fig:sketch}b we show how the energy macrostates of a ring with $N=1000$ spins are distributed with respect to their energy $\mathcal{H}$ and entropy $\log \brk{\Omega}$.
Each point in the figure corresponds to a specific energy macrostate, which can be  associated with a specific macrostate $\vec{m}_i = (S_{1;i}, S_{2;i})$.
They are colored according to the Boltzmann weights $\vec{\pi}(T_b)$ for $T_b=5$ which highlight a localization around the maximal entropy for a given energy.
In the thermodynamic limit the probability is concentrated around a specific value of the energy per particle, but also the density of macrostates around this value grows with $N$.

\begin{figure}
    \centering
    \includegraphics[width=\linewidth]{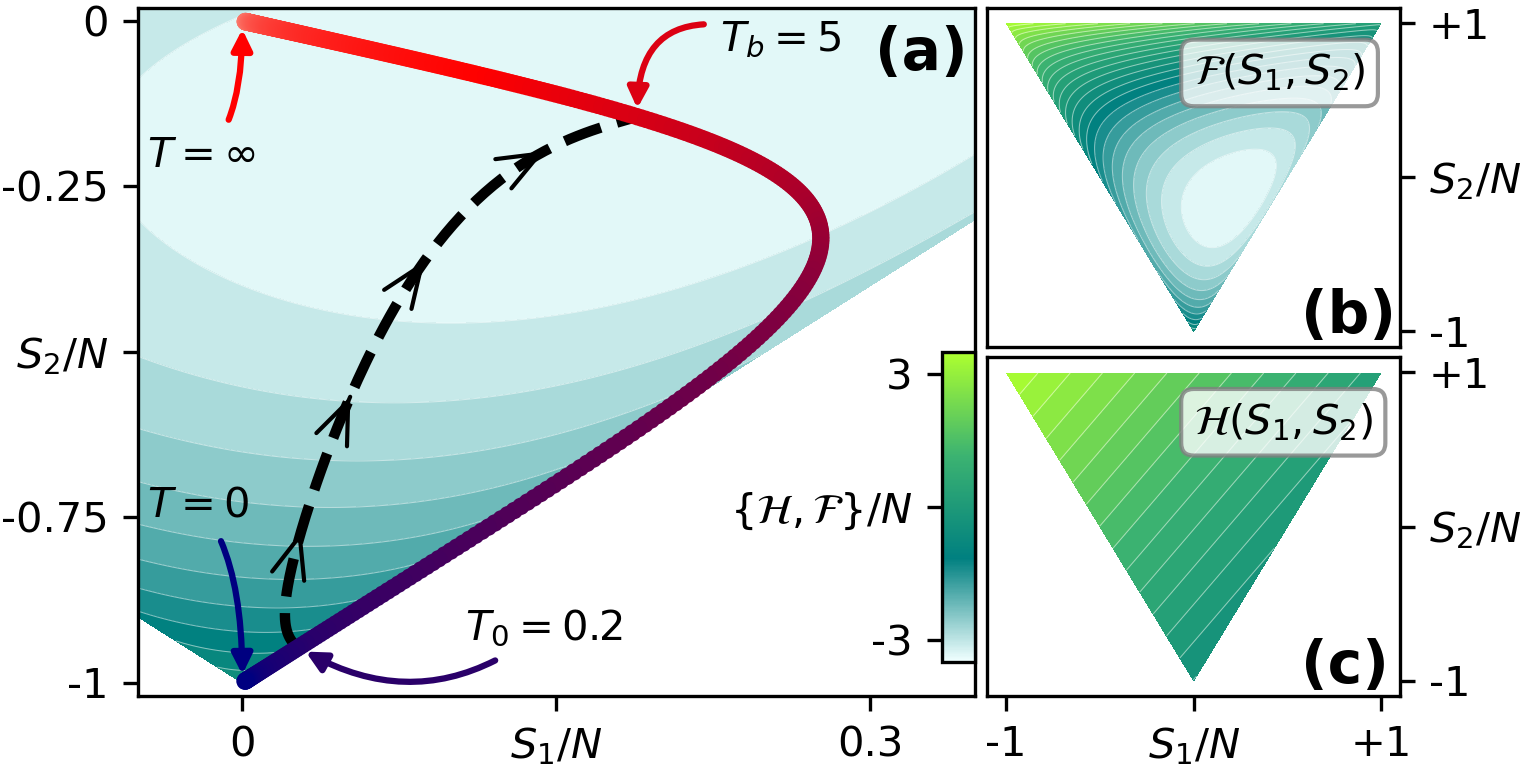}
    \caption{(a) Relaxation path from $T_0=0.2$ to $T_b=5$ (black) and equilibrium locus (blue-red gradient) projected on the $(S_1, S_2)$ configuration space.
    The relaxation process descends the gradient of the at bath temperature $\mathcal{F}(T_b)$ keeping the system consistently away from any equilibrium configuration.
    In panels (b) and (c) we plot the level curves of $\mathcal{F}$ and $\mathcal{H}$ over the whole configuration space.}
    \label{fig:relax_path}
\end{figure}

Our main result is that in the weak coupling limit the relaxation drives the system through configurations that are far from \textit{any} equilibrium distribution.
Setting the Boltzmann equilibrium $\vec{\pi}(T_0)$  as initial condition, Eq. \eqref{eq:mast_eq_solution} provides  the exact evolution of the system throughout the entire relaxation process.
The trajectory can be visualized by projecting the high-dimensional probability vector $\vec{p}(t)$ onto the $(S_1,S_2)$ configuration space, defined by the averages $\left< S_{\{1,2\}} \right>_{p}=\sum_i S_{\{1,2\};i} \cdot p_i(t)$ (see Fig. \ref{fig:relax_path}a for an example with $T_0=0.2$ and $T_b=5$).
Analogously, Boltzmann equilibrium distributions $\vec{\pi}(T)$ can be visualized on the same plane, resulting in a parametric curve of the temperature referred to as the \emph{equilibrium locus}.
A straightforward comparison between the two lines clearly shows how the relaxation process drives the system far from any equilibrium configuration.
Analyses at different sizes show that already at $N=20$ finite size effects become irrelevant (see Fig. \ref{fig:KL_en_fract}).
This suggests that the results obtained for $N=1000$ (corresponding to $\sim2.5\times10^5$ macrostates) adequately represent the thermodynamic limit.

To better understand the effect, in Fig. \ref{fig:relax_path}c we plot the energies $\mathcal{H}(S_1,S_2)$, demonstrating how {macrostates} that differ by a small energy density $\Delta E/N$ can still be far away in terms of transition steps of $S_1$ and $S_2$.
As a consequence, the relaxation bottleneck stems from the transitions among such macrostates, close in energy but far configuration-wise.
Indeed, examining the full set of macrostates for $N=10$ system size (Fig. \ref{fig:KL_en_fract}c) reveals that energy-adjacent states are far apart in terms of $S_1$ and $S_2$ transitions. 
Hence, even in the arbitrarily weak coupling case the system forcefully evolves across non-monotonous energy paths in order to reach any specific energy macrostate.
Remarkably, this mechanism also results in energy overshooting in the final stages of the relaxation, as shown in Fig. \ref{fig:KL_en_fract}b, which is reminiscent of the phenomenology observed in the Kovacs effect \cite{Kovacs1964,prados2014kovacs} and heated spin glass systems \cite{baity2019mpemba,bertin2013ageing}.

As system size $N$ increases, the topological structure of the boundary-driven dynamics between energy macrostates progressively enhances the non-equilibrium relaxation, which consolidates in the thermodynamic limit (Fig. \ref{fig:KL_en_fract}b-c).
Therefore, even in the thermodynamic limit,  where the equilibrium distribution concentrates in a single energy, the system is unable to relax quasistatically from one equilibrium state to another.
Instead, it is intrinsically forced to explore out-of-equilibrium distributions, so that the thermal quench has the same characteristic timescale as the boundary flips.

\begin{figure}
    \centering
    \includegraphics[width=\linewidth]{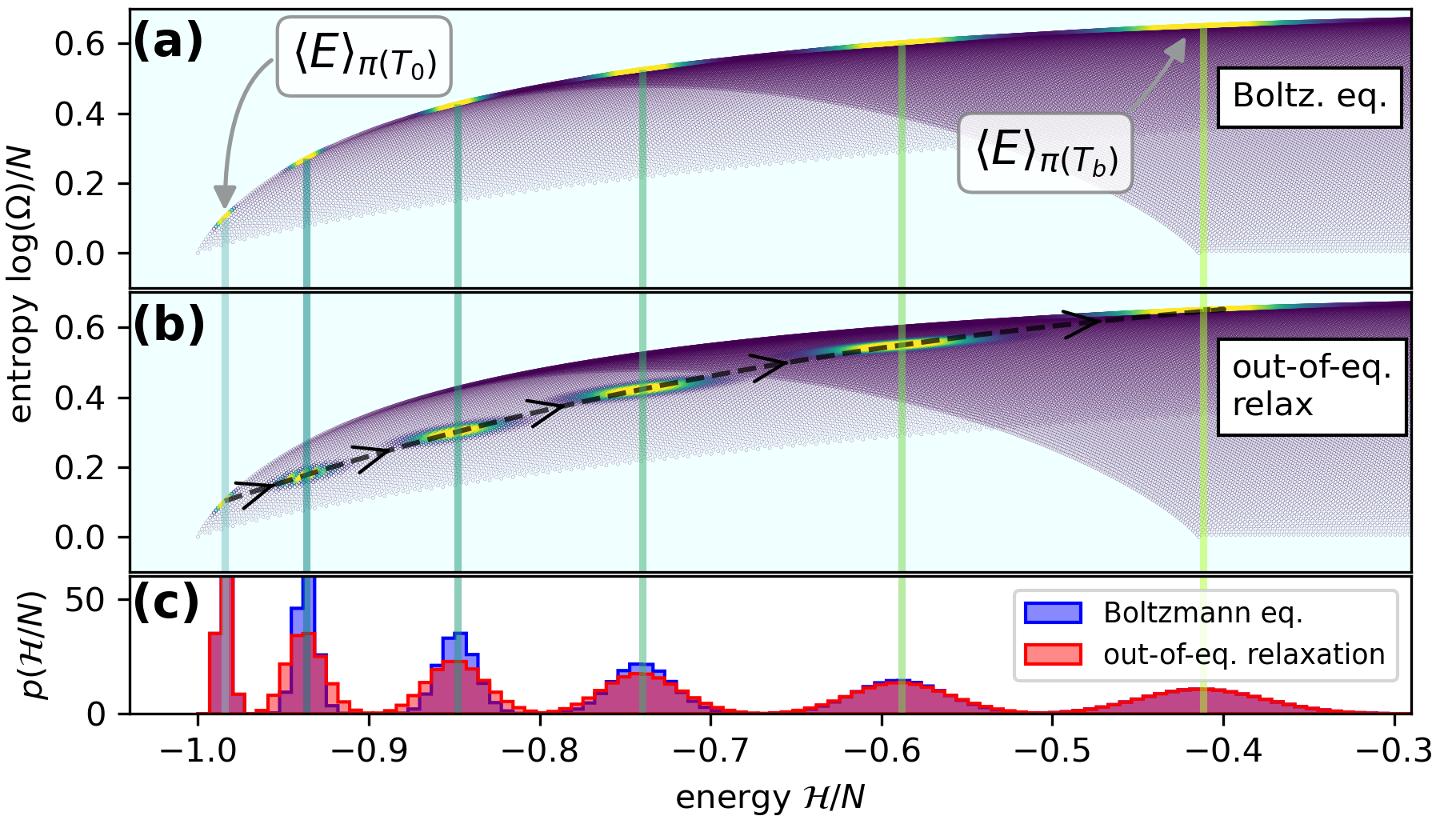}
    \caption{(a) The Boltzmann distribution for several temperatures (overlapped one on top of the other) in the energy-entropy plain introduced in Fig. \ref{fig:sketch}b (same colors as Fig. \ref{fig:sketch}b). The distributions are sharply peaked in the maximal entropy energy macrostates. (b) Instances of the out-of-equilibrium distributions during the relaxation path (black line) in the weak coupling limit. The mean energy of these instances are identical to the equilibrium distributions in (a), for comparison.
    (c) Projection on the energy axis shows how, in this system, the out-of-equilibrium energy distribution is in general wider than their equilibrium counterpart. A movie demonstrating the time evolution corresponding to this figure is given in the SM.}
    \label{fig:prob_path}
\end{figure}

Additional visualization that helps characterizing the relaxation more thoroughly is introduced in Fig. \ref{fig:prob_path}, where we fully visualize probability vectors as illustrated in Fig. \ref{fig:sketch}b.
In Fig. \ref{fig:prob_path}a we plot the equilibrium distribution $\vec{\pi}(T)$ of temperatures $T_0\leq T \leq T_b$, showing that they concentrate around the largest entropy configurations corresponding to the average energy $\left< E \right>_{\pi(T_0)}$ associated with each temperature. 
In Fig. \ref{fig:prob_path}b we plot snapshots of the distribution $\vec{p}(t)$ throughout the relaxation process, with the same average energy as the equilibrium distributions plotted in Fig. \ref{fig:prob_path}a (see the accompanying movie in the SM for the entire relaxation process).
From this comparison, it is clear how the configurations visited in the relaxation process do not correspond to any equilibrium distribution as predicted by Eq. \ref{eq:me_continuum}. The system maintains its configuration constantly at lower entropy in comparison with the corresponding equilibrium distribution, until it finally relaxes to $\vec{\pi}(T_b)$. Projecting this relaxation over the energy axis provides a histograms that allow to compare the energy distributions (Fig. \ref{fig:prob_path}c), generally showing a broader variance throughout the relaxation.

Summarizing, we have shown how macroscopic systems with discrete DoF can undergo far-from-equilibrium relaxations even in the arbitrarily weak coupling limit.
The relaxation occurs along a path that is heavily determined by the microscopic details of the dynamics (Eq. \ref{eq:me_continuum}), and in certain limits approaches the gradient descent of the LFE.
Such behaviour results from the discrete nature of DoF, which allows interactions among the DoF to force the system through non-monotonous energy trajectories.
The corrugated structure of the energy landscape consolidates with the system size, and survives the thermodynamic limit.
Our results pose a general warning on coarse-grained models that aim to describe out-of-equilbrium dynamics, such as those originally introduced by Landau \cite{Landau1965} and Ginzburg \cite{Landau2013} to address second order phase transitions in superfluids.
One common assumption in these models is that the system can be partitioned in mesoscopic cells that rapidly equilibrate at a local level.
This implies a \emph{continuum limit} \cite{blom2023thermodynamically} for some DoF, a procedure that we have proven to potentially disrupt microscopic mechanisms that are responsible for anomalous relaxation phenomena observed in macroscopic setups \cite{Jeng2006,chaddah2010overtaking,Hu2018,paper:hydrates,Kovacs1964}.
Our results explain how anomalous relaxation effects like the overshooting in energy (Fig. \ref{fig:KL_en_fract}b in Appendix B), the Mpemba effect  \cite{Teza2021_mpemba_boundary} and eigenvalue crossing \cite{teza2022eigenvalue} can still persist in the arbitrarily weak coupling limit for macroscopic systems.

% ----------------------- GT GOT HERE ------------------------

\begin{acknowledgments}
O. R. is supported by the Abramson Family Center for Young Scientists, the Israel Science Foundation Grant No. 950/19 and by the Minerva foundation.
G. T. acknowledges support by the Center for Statistical Mechanics at the Weizmann Institute of Science, the grant 662962 of the Simons foundation, the grants HALT and Hydrotronics of the EU Horizon 2020 program and the NSF-BSF grant 2020765.
We thank David Mukamel, Boaz Katz, Attilio L. Stella and Roi Holtzman for useful discussions.
\end{acknowledgments}

\bibliography{refs.bib}% Produces the bibliography via BibTeX.

\begin{appendix}

\renewcommand{\theequation}{A\arabic{equation}}
\setcounter{equation}{0}

\textit{Appendix A: Deriving the Continnum Limit Description}.
To obtain the master Eq. \eqref{eq:me_continuum} regulating the evolution in the weak coupling limit, we first rewrite master Eq. \eqref{eq:macro_master_eq} in terms of the transitions from state $\vec{m}_i$ to its neighbors $\vec{m}_i \pm \vec{\Delta}_{\alpha}$
\begin{align}\label{eq:sm_me}
    \begin{split}
        \partial_t p \left( \vec{m}_i, t \right) &= \frac{1}{2} \sum_{\alpha} [\mathbf{W}^{\text{weak}} \left(\vec{m}_i,  \vec{m}_i + \vec{\Delta}_{\alpha} \right) p \left( \vec{m}_i + \vec{\Delta}_{\alpha} , t \right) \\
        &+ \mathbf{W}^{\text{weak}} \left(\vec{m}_i,  \vec{m}_i - \vec{\Delta}_{\alpha} \right) p \left( \vec{m}_i - \vec{\Delta}_{\alpha} , t \right) \\
        &- \mathbf{W}^{\text{weak}}  \left(\vec{m}_i + \vec{\Delta}_{\alpha},  \vec{m}_i \right) p \left( \vec{m}_i, t \right) \\
        &- \mathbf{W}^{\text{weak}}  \left(\vec{m}_i - \vec{\Delta}_{\alpha},  \vec{m}_i  \right) p \left( \vec{m}_i, t \right)],
    \end{split}
\end{align}
where the $\frac{1}{2}$ factor is used to avoid double counting.
For convenience, we introduce the down/up transition rates 
\begin{align}
    \mathbf{W}^{\text{weak}} \left(\vec{m}_i,  \vec{m}_i + \vec{\Delta}_{\alpha}\right) & \equiv \mathbf{W}^{\text{weak}}_{\alpha; d} \left( \vec{m}_i + \vec{\Delta}_{\alpha} \right) \\
    \mathbf{W}^{\text{weak}} \left(\vec{m}_i,  \vec{m}_i - \vec{\Delta}_{\alpha}\right) & \equiv \mathbf{W}^{\text{weak}}_{\alpha;u} \left( \vec{m}_i - \vec{\Delta}_{\alpha}\right),
\end{align}
where $\mathbf{W}^{\text{weak}}_{\alpha; d/u}$ are the down/up transitions rates of macrostate $\vec{m}_i$ to its neighbors $\vec{m}_i \pm \vec{\Delta}_{\alpha}$.
Plugging this into Eq. \eqref{eq:sm_me} gives
\begin{align}\label{eq:sm_me_ud}
    \begin{split}
        \partial_t p \left( \vec{m}_i, t \right) &=
        \frac{1}{2} \sum_{\alpha} [ \mathbf{W}^{\text{weak}}_{\alpha;d} \left( \vec{m}_i + \vec{\Delta}_{\alpha} \right) p \left( \vec{m}_i + \vec{\Delta}_{\alpha} , t \right) \\
        &+ \mathbf{W}^{\text{weak}}_{\alpha;u} \left( \vec{m}_i - \vec{\Delta}_{\alpha}\right) p \left( \vec{m}_i - \vec{\Delta}_{\alpha} , t \right) \\
        &- \left( \mathbf{W}^{\text{weak}}_{\alpha;u} \left( \vec{m}_i\right) + \mathbf{W}^{\text{weak}}_{\alpha;d} \left( \vec{m}_i \right) \right) p \left( \vec{m}_i, t \right)].
    \end{split}
\end{align}
As $\vec{\Delta}_{\alpha}/\vec{m}_i\sim N^{-1}$, we can take the continuum limit and approximate Eq. \eqref{eq:sm_me_ud} as
\begin{equation}\label{eq:sm_continuum}
    \begin{aligned}
        \partial_t p \left( \vec{m}, t \right) = \frac{1}{2} \sum_{\alpha} \nabla_{\alpha}
        [ &\mathbf{W}^{\text{weak}}_{\alpha;d} \brk{ \vec{m} } p \left( \vec{m}, t \right) \\
        &-  \mathbf{W}^{\text{weak}}_{\alpha;u} \brk{ \vec{m} } p \left( \vec{m}, t \right) ].
    \end{aligned}
\end{equation}
Following the decomposition introduced in Eq. \eqref{eq:rates_decompose} we write 
\begin{equation}
    \Gamma_{ij} 
    = g  \brk{ \brk{ \mathcal{F}_j - \mathcal{F}_i }/ T_b  }
    = g  \brk{ \brk{ \mathcal{F} \brk{\vec{m}_j} - \mathcal{F} \brk{\vec{m}_i} }/ T_b  },
\end{equation}
for some even function $g$.
Recalling that transitions are only allowed to the nearest neighbors of $\vec{m}_i$ (which are $\vec{m}_j = \vec{m}_i \pm \vec{\Delta}_{\alpha}$) the free energy difference can be approximated as
\begin{equation}
    \mathcal{F}_j - \mathcal{F}_i 
    = \mathcal{F} \brk{\vec{m}_i \pm \vec{\Delta}_{\alpha}} - \mathcal{F} \brk{\vec{m}_i}
    \approx \pm \nabla_{\alpha} \mathcal{F} \brk{\vec{m}}.
\end{equation}
This allows to express the down/up transition rates as
\begin{align}
    \mathbf{W}^{\text{weak}}_{\alpha; d} \brk{ \vec{m} } 
    & \approx  g  \brk{\frac{1}{T_b} \nabla_{\alpha} \mathcal{F} \brk{ \vec{m} } } e^{ \frac{1}{2T_b} \nabla_{\alpha} \mathcal{F} \brk{ \vec{m} }} \label{eq:sm_transition_down}\\
    \mathbf{W}^{\text{weak}}_{\alpha;u} \brk{ \vec{m} }
    &  \approx  g  \brk{ \frac{1}{T_b} \nabla_{\alpha} \mathcal{F}  \brk{ \vec{m} } } e^{ -\frac{1}{2 T_b} \nabla_{\alpha} \mathcal{F} \brk{ \vec{m} }} \label{eq:sm_transition_up},
\end{align}
where we used the fact the $g$ is even.
Plugging Eqs. \eqref{eq:sm_transition_down},\eqref{eq:sm_transition_up} into Eq. \eqref{eq:sm_continuum} one recovers Eq. \eqref{eq:me_continuum}.

In the most general case $\Gamma_{ij}$ does not depend only on the free energies differences.
In this case, we can approximate it in the continuum limit as
\begin{equation} \label{eq:sm_non_fe_diff}
    \Gamma_{ij} 
    = \tilde{g}  \brk{ \mathcal{F} \brk{\vec{m}_i} , \mathcal{F} \brk{\vec{m}_i \pm \vec{\Delta}_{\alpha}}}
    \approx \tilde{g}  \brk{ \mathcal{F} \brk{\vec{m}} , \mathcal{F} \brk{\vec{m}}},
\end{equation}
with some even function $\tilde{g}$.
As this term does not vanish when $\Gamma_{ij}$ depends on both free energies, in this case it is sufficient to take the zeroth order approximation.
In addition, the next term scales like $\sim 1/N\ll 1$, which is negligible in the thermodynamic limit.
Plugging Eq. \eqref{eq:sm_non_fe_diff} into Eq. \eqref{eq:sm_continuum} yields 
\begin{equation} \label{eq:sm_me_non_fe_diff}
    \begin{aligned}
        \partial_t p \brk{ \vec{m} , t}
        = 
        \sum_{\alpha} \nabla_{\alpha} [ &  \tilde{g} \brk{ \mathcal{F} \brk{\vec{m}},  \mathcal{F} \brk{\vec{m}} }  \\
        & \times \sinh \brk{ \frac{1}{2 T_b} \nabla_\alpha \mathcal{F} \brk{ \vec{m} } }  p \brk{ \vec{m} , t}].
    \end{aligned}
\end{equation}
Also this dynamics contains the same features of Eq. \ref{eq:me_continuum} that prevent the system from relaxing quasistatically, implying the same holds also in the most general case where Eq. \ref{eq:sm_non_fe_diff} holds.

\textit{Glauber rates -- }
Glauber dynamics \cite{Glauber1963} assumes transition rates with a shape like those in Eq. \ref{eq:rate_mat_weak} with 
$\Gamma_{ij} = \frac{1}{2}\textrm{sech } \frac{\mathcal{F}(\vec{m}_j) - \mathcal{F}(\vec{m}_i) }{2 T_b}$, implying that $\mathbf{W}^{\textrm{weak}}$
solely depend on free energy differences.
Following the derivation in the manuscript, one obtains an analogous dynamics equation to Eq. \eqref{eq:me_continuum} with
\begin{equation}
    g_{GL} \brk{ \frac{\nabla_\alpha \mathcal{F} \brk{\vec{m}}}{T_b}  }
    = \frac{1}{2}\textrm{sech } \frac{\nabla_\alpha \mathcal{F}\brk{\vec{m}} }{2 T_b}\ ,
\end{equation}
showing that Glauber dynamics also yield far from equilibrium relaxation in the weak coupling limit (here $g_{GL}$ denotes the even function specific to Glauber dynamics).

\begin{figure}
    \centering
    \includegraphics[width=\linewidth]{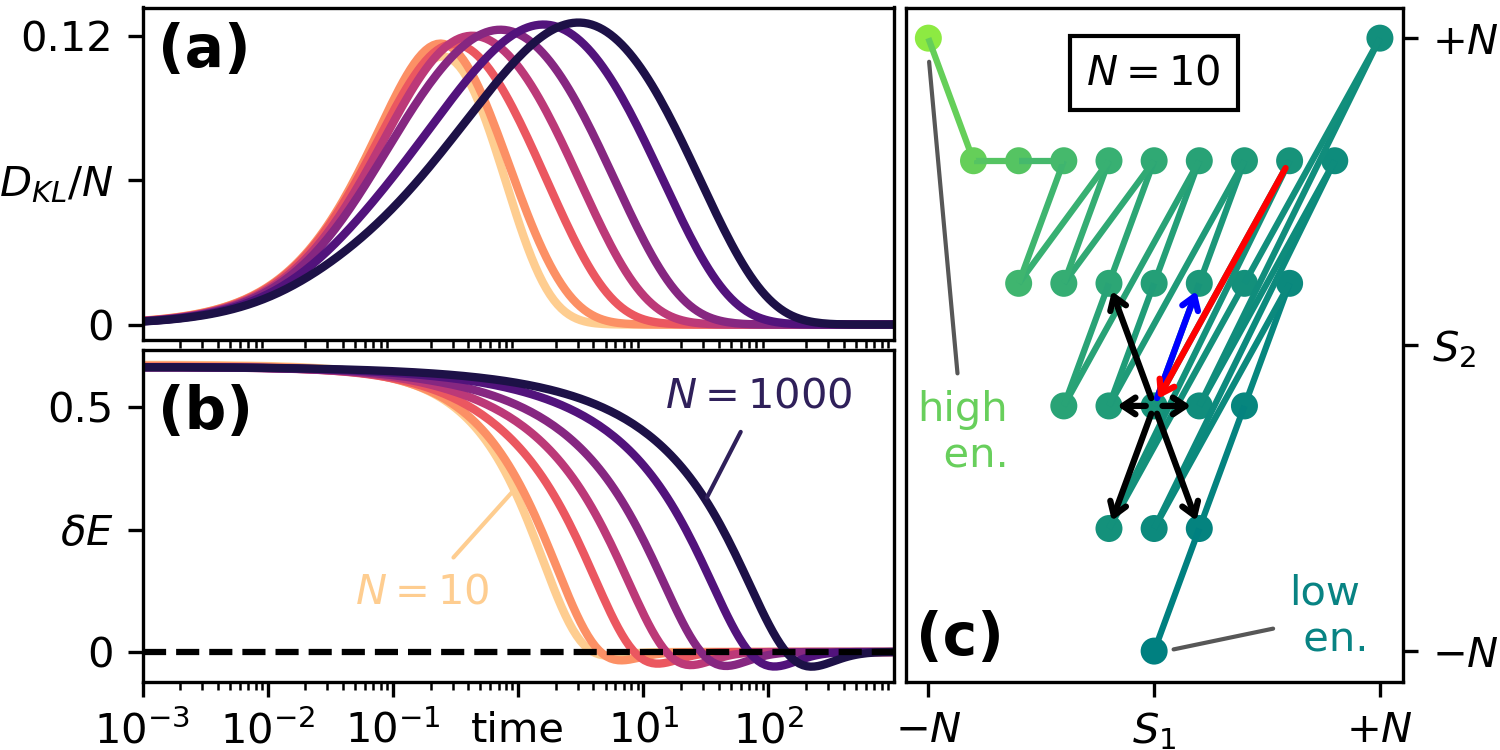}
    \caption{(a) Evolution of the KL divergence between $p(t)$ and the Boltzmann equilibrium at the corresponding average energy, for different sizes $N$. It shows that the system quickly departs from the initial $T_0$ equilibrium before slowly approaching the final one at $T_b$ (the time axis is in log-scale).
    (b) Evolution of the relative energy difference $\delta E=(E_b-\left< E \right>_p)/E_b$, exhibiting an overshoot of the target energy of the system at bath temperature $E_b$.
    (c) Corrugated nature of the energy landscape. Adjacent energy macrostates are connected with a segment, defining the increasing energy path (colors follow those of Fig. \ref{fig:relax_path}).
    The arrows indicate the possible connections ($\Delta S_1=\pm2$ and $\Delta S_2 = \{0,\pm,4\}$), for an exemplificative energy macrostates. Highlighted in blue is the possible transition to the next higher in energy state, while the red arrow shows how the preceding one is not directly connected.}
    \label{fig:KL_en_fract}
\end{figure}

\textit{Metropolis algorithm --}
The Metropolis dynamics is another example of transition rates that depend solely on the free energy difference
\begin{equation}
    \begin{aligned}
       \mathbf{W}_M \brk{\vec{m}_i, \vec{m}_j} 
       & = 
        \begin{cases}
            1  & \mathcal{F} \brk{\vec{m}_i} \le \mathcal{F} \brk{\vec{m}_j} \\
            e^{ -  \frac{1}{T_b} \left(\mathcal{F} \brk{\vec{m}_i} - \mathcal{F} \brk{\vec{m}_j}\right)} & \mathcal{F} \brk{\vec{m}_i} \ge \mathcal{F} \brk{\vec{m}_j}
        \end{cases} \\ 
        & = e^{ -  \frac{1}{2 T_b} \left(\mathcal{F} \brk{\vec{m}_i} - \mathcal{F} \brk{\vec{m}_j} + \abs{\mathcal{F} \brk{\vec{m}_i} - \mathcal{F} \brk{\vec{m}_j}}\right)} 
    \end{aligned}
\end{equation}
In this case the symmetric matrix take the form 
\begin{equation}
        \Gamma_{ij} 
        = g_{M} \left( \frac{\mathcal{F} \brk{\vec{m}_i} - \mathcal{F} \brk{\vec{m}_j}}{T_b} \right) 
        = e^{ - \frac{1 }{2 T_b} \abs{\mathcal{F} \brk{\vec{m}_i} - \mathcal{F} \brk{\vec{m}_j}}},
\end{equation}
where we denote $g_{M}$ as the even function corresponding to Metropolis dynamics.
Following Eq. \eqref{eq:me_continuum} we write
\begin{equation}
    g_{M} \brk{ \frac{1}{T_b} \nabla_\alpha \mathcal{F} \brk{\vec{m}} }
    = e^{ - \frac{ 1 }{2 T_b} \abs{\nabla_\alpha \mathcal{F} \brk{\vec{m}} }},
\end{equation}
showing that also Metropolis dynamics deviate from a quasi static process in the weak coupling limit.

\textit{Retrieving gradient descent dynamics ---}
The intuitive picture where the dynamics of Eq. \eqref{eq:me_continuum} follows the LFE gradient is retrieved when the LFE gradients are much smaller than the bath temperature $\nabla_\alpha \mathcal{F} \brk{ \vec{m}} \ll T_b$. In this case, the dynamics simplifies to
\begin{equation} \label{eq:grad_descent}
    \partial_t p \brk{ \vec{m}, t } 
    = \frac{\Gamma_0 }{2 T_b} \sum_{\alpha} \nabla_{\alpha} \left[ \brk{\nabla_{\alpha} \mathcal{F} \brk{ \vec{m}}}  p \brk{ \vec{m}, t } \right],
\end{equation}
where we introduced the constant $\Gamma_0 = g \brk{0} $.
Eq. \eqref{eq:grad_descent} has clearly the shape of a Fokker-Planck equation, with the free energy $\mathcal{F}$ acting as a potential defining the landscape for the evolution, which occurs along its gradient descent.
Hence, even in this limit the relaxation does not follow the quasistatic limit and the system explores configurations that are away from any equilibrium distribution, enabling the emergence of anomalous relaxation effects.

\renewcommand{\theequation}{B\arabic{equation}}
\setcounter{equation}{0}

\textit{Appendix B: Finite size analysis.}
Fig. \ref{fig:KL_en_fract} provides complementary support of analysis of the far from equilibrium relaxation in the weak coupling limit.
First, in Fig. \ref{fig:KL_en_fract}a, we show numerical evidence that the effect persists as the number of spins $N$ increases.
This is highlighted by plotting the Kullback-Leibler (KL) divergence \cite{Kullback1951} between the relaxation probability $\vec{p}\brk{t}$ and the equilibrium with corresponding average energy, denoted by $\vec{\pi}\brk{T\brk{t}}$
\begin{equation}\label{eq:DKL}
    D_{KL}(t)=\sum_i \pi_i \brk{T\brk{t}} \log \left[ p_i \brk{t}/\pi_i\brk{T\brk{t}} \right].
\end{equation}
The KL divergence exhibits a marked peak, indicating a departure from the equilibrium locus which is eventually reached on a longer time-scale, during the final stages of the relaxation \footnote{Note that other comparisons to equilibrium distribution are possible, and our choice is natural but not the only way to measure distance from equilibrium.}.
Analyzing a wide range of system sizes $N=\{10,20,\dots,1000\}$ suggests that this characterization survives the thermodynamic limit, ruling out the possibility that the far-from-equilibrium relaxation is due to finite-size effects. 

Fig. \ref{fig:KL_en_fract}c shows the full set of {macrostates} for a system size of $N=10$: the macrostates are connected according to their energy, from lowest to highest, exhibiting non-monotonic behavior on this plane.
Marking the connectivity between the macrostates, in terms of $S_1$ and $S_2$ transitions, with arrows for a specific macrostate clarifies the discrepancy between possible transitions and the following/preceding energy macrostate, forcing non-monotonous energy paths. 
Lastly, Fig. \ref{fig:KL_en_fract}b illustrates how energy paths overshoot the final energy. 

\begin{figure}[b]
    \centering
    \includegraphics[width=\linewidth]{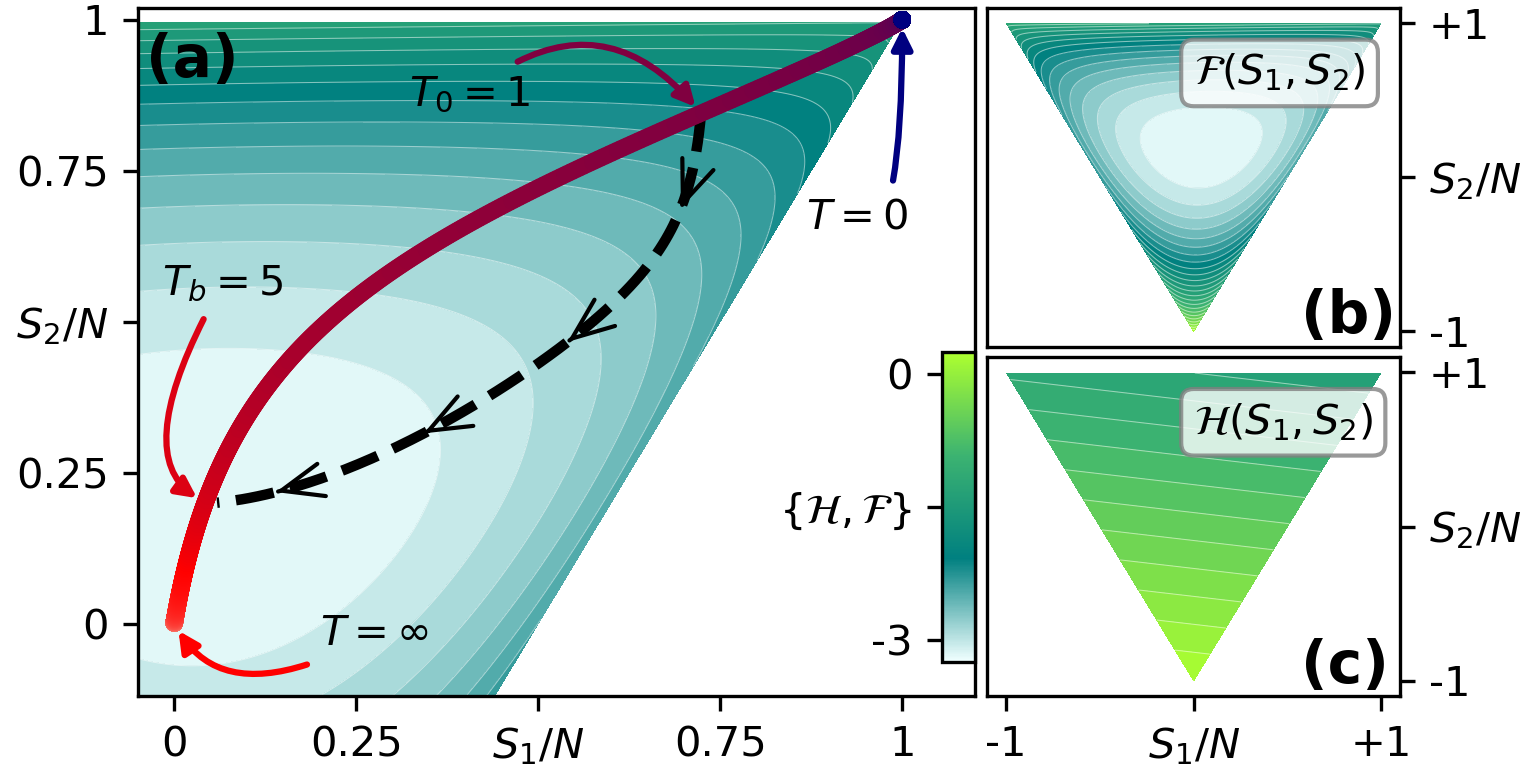}
    \caption{Analogous of Fig. \ref{fig:relax_path} for an Ising ferromagnet ($J=+1$ in Eq. \ref{eq:ham_AF}). (a) We report the relaxation path (black)  of a thermal quench from initial temperature $T_0=1$ to bath temperature $T_b=5$ with external magnetic field $H=\sqrt{2}/10$, which keeps consistently away from any equilibrium configuration (gradient line).
    In panels (b) and (c) we plot the level curves of $\mathcal{F}$ and $\mathcal{H}$ over the whole configuration space.}
    \label{fig:S1_S2_ferro}
\end{figure}

\end{appendix}

\end{document}